\documentclass{elsart5p}
\usepackage{graphics}
\usepackage{graphicx}
\usepackage{epsfig}
\usepackage{amssymb}
\usepackage{amsmath}

\renewcommand{\pol}[1]{\mathaccent"017E{#1}}
\newcommand{\reactiona}{\mbox{$pd \rightarrow{}^3 \textrm{He}\,\eta$}}
\newcommand{\reactionb}{\mbox{$dp \rightarrow{}^3 \textrm{He}\,\eta$}}
\newcommand{\reactionc}{\mbox{$\pol{d}p \rightarrow{}^3 \textrm{He}\,\eta$}}

\begin{document}
\begin{frontmatter}

\title{Absence of spin dependence in the final state interaction of the $\pol{d}p\rightarrow{}^3$He$\eta$  reaction}

\author[Muenster]{M.~Papenbrock\corauthref{cor1}}
\ead{michaelp@uni-muenster.de} \corauth[cor1]{Corresponding author.},
\author[Gatchina]{S.~Barsov},
\author[Muenster]{I.~Burmeister}$^{,1}$\thanks{Present address: Fakult\"{a}t Physik, Technische Universit\"{a}t Dortmund, D-22441 Dortmund, Germany},
\author[IKP,Tbilisi]{D.~Chiladze},
\author[IKP,Dubna]{S.~Dymov},
\author[Muenster]{C.~Fritzsch},
\author[IKP]{R.~Gebel},
\author[Muenster]{P.~Goslawski}$^{,2}$\thanks{Present address: Helmholtz-Zentrum Berlin f\"ur Materialien und Energie GmbH, Elektronen-Speicherring BESSY II, Germany},
\author[IKP]{M.~Hartmann},
\author[IKP]{A.~Kacharava},
\author[Basel]{I.~Keshelashvili}$^{,3}$\thanks{Present address: Institut f\"ur Kernphysik and J\"{u}lich Centre for Hadron Physics, Forschungszentrum J\"ulich, D-52425 J\"ulich, Germany},
\author[Muenster]{A.~Khoukaz},
\author[Cracow]{P.~Kulessa},
\author[Dubna]{A.~Kulikov},
\author[IKP]{B.~Lorentz},
\author[IKP,Tbilisi]{D.~Mchedlishvili},
\author[Muenster]{T.~Mersmann},
\author[IKP]{S.~Merzliakov},
\author[Muenster]{M.~Mielke},
\author[Gatchina,IKP]{S.~Mikirtychiants}
\author[IKP]{H.~Ohm},
\author[IKP]{D.~Prasuhn},
\author[IKP]{F.~Rathmann},
\author[Muenster]{T.~Rausmann}$^{,4}$\thanks{Present address: T\"UV Nord SysTec Gmbh \& Co. KG, D-22525 Hamburg},
\author[IKP]{V.~Serdyuk},
\author[IKP]{H.~Str\"{o}her},
\author[Muenster]{A.~T\"{a}schner},
\author[Rossendorf,Moscow]{S.~Trusov},
\author[Gatchina,Bonn]{Y.~Valdau},
\author[UCL]{C.~Wilkin}

\address[Muenster]{Institut f\"ur Kernphysik, Westf\"alische
Wilhelms-Universit\"at M\"unster, D-48149 M\"unster, Germany}
\address[Gatchina]{High Energy Physics Department, Petersburg Nuclear Physics Institute, RU-188350 Gatchina, Russia}
\address[IKP]{Institut f\"ur Kernphysik and J\"{u}lich Centre for Hadron Physics, Forschungszentrum J\"ulich, D-52425 J\"ulich, Germany}
\address[Tbilisi]{High Energy Physics Institute, Tbilisi State University, GE-0186 Tbilisi, Georgia}
\address[Dubna]{Laboratory of Nuclear Problems, JINR, RU-141980 Dubna, Russia}
\address[Basel]{Department of Physics, University of Basel, CH-4056 Basel, Switzerland}
\address[Cracow]{H.~Niewodniczanski Institute of Nuclear Physics PAN, PL-31342 Cracow, Poland}
\address[Rossendorf]{Institut f\"ur Kern- und Hadronenphysik, Forschungszentrum Rossendorf, D-01314 Dresden, Germany}
\address[Moscow]{Skobeltsyn Institute of Nuclear Physics, Lomonosov Moscow State University, RU-119991 Moscow, Russia}
\address[Bonn]{Helmholtz-Institut f\"ur Strahlen- und Kernphysik, Universit\"at Bonn, D-53115 Bonn, Germany}
\address[UCL]{Physics and Astronomy Department, UCL, London WC1E 6BT, UK}

\begin{abstract}
The deuteron tensor analysing power $t_{20}$ of the \reactionc\
reaction has been measured at the COSY-ANKE facility in small
steps in excess energy $Q$ up to $Q=11$~MeV. Despite the
square of the production amplitude varying by over a factor of
five through this range, $t_{20}$ shows little or no energy
dependence. This is evidence that the final state interaction
causing the energy variation is not influenced by the spin
configuration in the entrance channel. The weak angular
dependence observed for $t_{20}$ provides useful insight into
the amplitude structure near threshold.
\end{abstract}

\end{frontmatter}

%

It has been known for many years that measurements of the total cross section
for the \reactiona\ or \reactionb\ reaction show very anomalous results near
threshold~\cite{BER1988,MAY1996,MER2007,SMY2007}. In terms of the excess
energy $Q=W-M_{^3{}\textrm{He}}-M_{\eta}$, where $W$ is the total
centre-of-mass (c.m.) energy, the cross section jumps to its plateau value
already for $Q<1$~MeV. It was suggested that this effect is due to a strong
final state interaction (FSI) between the $\eta$ and the $^3$He that might
even lead to a quasi-bound $\eta$-nuclear state~\cite{WIL1993}. Fits to the
most detailed data set have been made that show a pole in the $\eta\,^3$He
elastic amplitude for $|Q|<0.5$~MeV~\cite{MER2007}. Support for this pole
assumption also comes from the study of the variation of the angular
dependence of the cross section with $Q$~\cite{WIL2007}.

If the threshold behaviour is due to an $\eta\,^3$He FSI then this should
manifest itself in broadly similar ways for different entrance channels. The
new photoproduction data on $\gamma{}^3\textrm{He} \to \eta{}^3\textrm{He}$
show a steep rise in the first 4~MeV excess energy bin above
threshold~\cite{PHE2012}, though their resolution or statistics were not
sufficient to determine the pole position with high precision.

In the \reactionb\ reaction it is possible to access the $s$-wave
$\eta\,^3$He system from either the total spin $S=\frac{3}{2}$ or the
$S=\frac{1}{2}$ initial states and the differences will influence
measurements of the deuteron tensor analysing power $t_{20}$ of the
\reactionc\ reaction. The pure $s$-wave FSI hypothesis would require that the
value of $t_{20}$ should remain constant as a function of energy even though
the corresponding unpolarised cross section varies. This approach to the
comparison of the effects of the entrance channel on the FSI is very
appealing because the detection system is independent of the deuteron beam
polarisation and many of the systematic effects cancel.

The only $t_{20}$ measurement in this reaction was carried out at four
energies close to threshold at Saclay~\cite{BER1988}, reaching up to $Q
\approx 5$~MeV. Over the limited energy range their values of $t_{20}$ were
consistent with being constant, though the results were not compelling. We
have therefore remeasured this observable in greater detail and over a wider
energy range.

The ANKE magnetic spectrometer~\cite{BAR1997} is located at an internal
target station of the COoler SYnchrotron (COSY) of the Forschungszentrum
J\"{u}lich. Only the forward detection system was used in the present
experiment and this has full geometric acceptance for the \reactionb\
reaction at the energies investigated up to $Q=11$~MeV. With an unpolarised
deuteron beam incident on a hydrogen cluster-jet target~\cite{KHO1999}, this
facility was already used to measure the unpolarised cross sections at 192
values of the excess energy~\cite{MER2007}. Since the same system was also
used when repeating the experiment with a polarised deuteron beam, much of
the experimental discussion will here be focussed on the polarisation
effects.

Polarised $D^{-}$ ions are produced by colliding atomic deuterium with a
caesium beam. A series of permanent sextupole magnets and radio frequency
transition units is then used to enhance further the occupation numbers of
the desired polarisation state~\cite{FEL2009}. The negatively charged ions
are then pre-accelerated to the injection energy but, before entering the
COSY ring, they are stripped of their electrons when passing through a carbon
foil.

Three different polarisation modes plus one unpolarised mode for polarimetry
were provided by the ion source. The modes were arranged to provide different
vector ($P_z$) and tensor ($P_{zz}$) polarisations, alternating with each
beam injection, in order to be able to compare them and investigate
systematics. The polarisations refer to an axis that is perpendicular to the
accelerator plane, being conventionally labeled by $z$ in the source frame.
Though the nominal values of the beam polarisations are shown in
Table~\ref{polarisationtable}, these are idealised numbers and the actual
ones are better determined in the experiment itself. The procedure for doing
this will be discussed later and the mean values of the measured $P_{zz}$ are
shown in the table. It is seen here that the tensor polarisation for mode~3
was about a third of that of mode~2 so that it is clear that one or more of
the hyperfine transitions in this mode were not operating as desired. The
results presented in this letter are therefore based upon the data taken with
modes 1 and 2 plus the unpolarised mode.

\begin{table}[h]
\begin{center}
\begin{tabular}{ccccc}
Mode &\phantom{1} $P_{z}^{\rm ideal}$\phantom{1} & $P_{zz}^{\rm
ideal}$&$P_{z}^{\rm LEP}$&$P_{zz}^{\rm ANKE}$\\
\hline
1 & +1/3 & --1 & $+0.244\pm0.032$\phantom{0} & $-0.62\pm 0.05$\\
2 & --1 & +1 & $-0.707\pm 0.026$\phantom{0} &$+0.67\pm 0.05$\\
3 & +1 & +1 & $+0.601\pm0.027$\phantom{0} & $+0.22\pm0.05$\\
\end{tabular}
\end{center}
\caption{Nominal values of the vector and tensor polarisations of the
deuterons provided by the ion source for the three polarisation modes. Also
shown are the average tensor polarisations $P_{zz}^{\rm ANKE}$ measured
through the ramp using the $\pol{d}p\to\{pp\}_sn$ reaction. The values quoted
for $P_{z}^{\rm LEP}$ were measured at the injection energy of 75.6~MeV with
the low energy polarimeter~\cite{FEL2009} for a sample of deuterons.}
\label{polarisationtable}
\end{table}

Also shown in Table~\ref{polarisationtable} are the values of the vector
polarisation measured with the low energy polarimeter at the 75.6~MeV
injection energy. Though the deuterons are unlikely to lose any polarisation
through the acceleration, we cannot guarantee that the source was completely
stable such as to make these values valid throughout the whole experiment.

The beam was continuously ramped from a deuteron beam momentum of
3.118~GeV/$c$ up to 3.185~GeV/$c$ over a 300 second cycle. While the beam
momentum is only known with an uncertainty of $\Delta p /p \approx 10^{-3}$
from macroscopic measurements, the study of the $^3$He momentum ellipse from
the \reactionb\ reaction allows the production threshold to be determined
with much better precision~\cite{MER2007}. The momentum range then
corresponds to an excess energy of $-5$~MeV to $+11$~MeV. The below-threshold
data were taken for the evaluation of the background.

Several hardware triggers were installed to filter out unwanted background
events and thus minimise the dead time. A high threshold trigger, which was
designed mainly for the primary reaction, selected events with a high energy
loss in the hodoscopes of the forward detection system. This eliminated most
of the charge-one protons and deuterons.  A second trigger, which looked for
events where there was an energy loss in one of the first two layers of the
forward hodoscopes, was prescaled by a factor of 1024. The associated data
were used for luminosity and polarisation determinations. A final trigger
stored information on the experimental conditions (e.g. beam intensity and
event rates) which, for instance, are necessary to estimate the dead time of
the data acquisition.

\begin{figure}[h]
\centering
\includegraphics[width=1.0\linewidth]{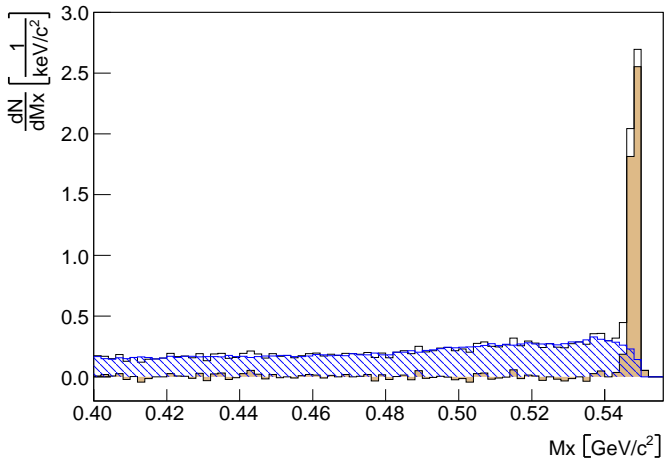}
\caption{Missing-mass distribution of the
$\pol{d}p\to{}^3\textrm{He}\,X$ at an excess energy of 3~MeV
with respect to the $\eta$ threshold (black line). Data taken
below the threshold (blue line) were transformed to this energy
and fitted to the data in the 0.4~GeV/$c^2$ to 0.52~GeV/$c^2$
missing-mass region. After subtracting the background, only a
clean (brown shaded) missing-mass peak remains.}
\label{BGCorrection}
\end{figure}

The $^3 \text{He}$ nuclei were detected in the ANKE forward detection system
and the $\eta$ mesons reconstructed through the missing-mass peak. The shape
of the missing-mass spectra depends mainly on the characteristics of the
spectrometer and changes little over small steps in beam energy. Analysing
the subthreshold data as if they were taken above the $\eta$ production
energy, they can be used to provide an accurate description of the background
below the $\eta$ missing-mass peaks. This procedure, which was already used
successfully for the unpolarised data~\cite{MER2007}, is illustrated in
Fig.~\ref{BGCorrection}. This technique was applied for all the energy bins
and polarisation modes used in the data analysis.

The tensor polarisation of the deuteron beam was determined from the study of
the $\pol{d}p \rightarrow \{pp\}_sn$ reaction that was measured
simultaneously through the ramp. For this purpose, events with two protons in
the forward detection system were analysed. If the two proton system
$\{pp\}_s$ is at low excitation energy, typically $E_{pp}<3$~MeV, the tensor
analysing power signal is very strong and can be predicted with high accuracy
in the required energy range~\cite{CAR1991,MCH2013}. The vector analysing
power is, as expected~\cite{BUG1987}, consistent with zero~\cite{CHI2009}.
The tensor polarisation could therefore be determined at the same energies as
those used for the main reaction. However, there are no depolarising
resonances for deuterons in the COSY energy range and so the values of the
polarisations could also be checked at the standard beam energy of
1.2~GeV~\cite{CHI2009} in each cycle. Any deviations observed here were very
small and would affect the mean value of $t_{20}$ by less than $0.02$ while
not influencing the energy or angular dependence of the observable.

For a two-body or quasi-two-body reaction, the ratio of the numbers of
polarised to non-polarised events becomes
\begin{eqnarray}
\nonumber
\frac{N^{\uparrow}}{N^{0}}  &=&
C_n\left(1 + \sqrt{3}P_{z} i t_{11} (\theta) \cos\phi\right.\\
&&\hspace{1cm}\left. - \half P_{zz}\left[t_{20}(\theta)/\sqrt{2}
+\sqrt{3}t_{22}(\theta)\cos2\phi\right]\right),
\label{pzzformula}
\end{eqnarray}
where $\phi$ is the azimuthal angle, $it_{11}$ the (spherical) vector
analysing power, and $t_{20}$ and $t_{22}$ two components of the tensor
analysing power~\cite{OHL1972}.

Since the $t_{20}$ tensor signal is the strongest and has no dependence on
the azimuthal angle, the relative luminosity of the polarised to unpolarised
modes $C_n$ must be determined accurately. Though this could be estimated
from the beam current transformer (BCT) signal, the most precise measurement
is found by counting simultaneously the numbers of spectator protons
registered in the forward detection system with a Fermi momentum $p_{\rm
spec} \leq 60$~MeV/$c$. These rates are largely unaffected by the beam
polarisation~\cite{MCH2013} and we did not find any significant change in the
values of $C_n$ when reducing the cut to 40~MeV/$c$. Not only does this
method account for the different luminosities but, since they were obtained
using the same hardware trigger, the dead-time effects are also included.
Furthermore, the geometrical acceptances are identical for the polarised and
unpolarised modes.

Equation~(\ref{pzzformula}) is used with the $\pol{d}p \rightarrow \{pp\}_sn$
data to determine the values of $P_{zz}$ for each mode in terms of the known
values of $t_{20}$ and $t_{22}$~\cite{MCH2013}. It is also used to extract
values of $t_{20}$ for the \reactionc\ reaction, using precisely these values
of $P_{zz}$. However, the $\pol{d}p \rightarrow \{pp\}_sn$ reaction is not
sensitive to the vector polarisation of the beam and alternative solutions,
such as the study of the quasi-free $\pol{n}p\to d\pi^0$~\cite{CHI2006}, are
strongly affected by the background conditions. We are therefore forced to
use the values of $P_{z}^{\rm LEP}$ quoted in Table~\ref{polarisationtable}
when extracting $it_{11}$, though these polarisations were not measured
throughout the whole run.

Unfortunately, it was found during the analysis that, due to a defective
scintillation counter in the third forward hodoscope layer, there was not a
complete angular acceptance for the \reactionc\ reaction across the full
range of excess energies. The data were therefore grouped into 10 bins in
$\cos\theta$ and 4 bins in $\phi$ in such a way that $\int \cos2\phi\, \text{d}\phi=
0$ for each $\cos\theta$ bin. This eliminates any possible contributions
arising from $t_{22}$.

At the low excess energies studied in this experiment, the vector analysing
power signal is proportional to $\sin\theta\cos\phi$. Using only those
angular regions with full azimuthal acceptance allowed the values of
$it_{11}$ to be determined from the $\cos\phi$ dependence. The results of the
fits confirmed that $|it_{11}|\lesssim 0.04$ for $Q\leqslant 10$~MeV. One
slight caveat here is that this analysis does rely on the values of
$P_{z}^{\rm LEP}$ quoted in Table~\ref{polarisationtable} and these were not
determined for all the data-taking runs.

\begin{figure}[h]
\includegraphics[width=1.0\linewidth]{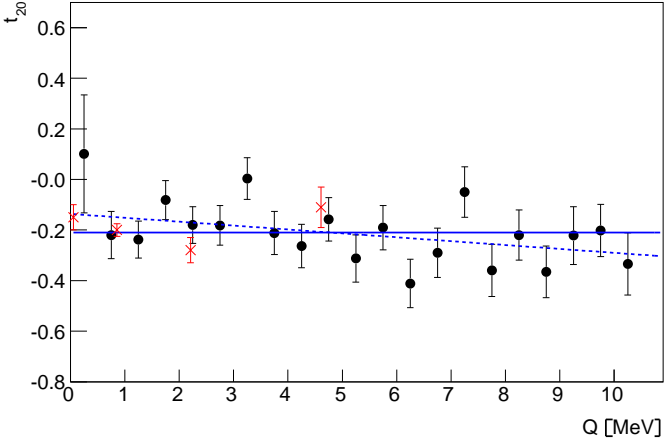}
\caption{The tensor analysing power $t_{20}$ (black circles) measured for the
\reactionc\ reaction in $0.5$~MeV bins in the excess energy $Q$. The solid
and dashed lines are, respectively, constant and linear fits to these data.
Also shown are the four points (red crosses) measured at
Saclay~\cite{BER1988}.} \label{t20combined}
\end{figure}

Having extracted the vector polarisation asymmetries for all $\theta$
and $Q$, $t_{20}$ could then be determined from Eq.~(\ref{pzzformula}) over
the whole $\cos\theta$ range for each energy. The values obtained for
polarisation modes 1 and 2 are consistent within $2 \sigma$ and the results
found by averaging over the polarisation modes and all $^3$He c.m.\ angles
are shown in Fig.~\ref{t20combined}. Since the data were taken over a
continuous ramp, the binning in $Q$ is somewhat arbitrary and the $0.5$~MeV
used here was chosen to match the statistics. The results are largely in
agreement with those measured at Saclay~\cite{BER1988}.

The ANKE data shown in Fig.~\ref{t20combined} are consistent with a constant
value of $t_{20}=-0.21\pm 0.02\pm 0.05$, which has a reduced $\chi^2$ of
1.34. Here the first error is statistical and the second systematic. The
latter arises mainly from uncertainties in the beam polarisation, the
relative flux normalisation, the geometry, and the contribution from
$it_{11}$.

Allowing $t_{20}$ to have a linear dependence on $Q$ leads to a marginal
improvement in the description of the data with
\begin{equation}
\label{bestfit}
t_{20}=(-0.14\pm0.04)+ (-0.02\pm0.01)Q,
\end{equation}
where $\chi^2/\textrm{NDF}=1.17$ and here and in other fits $Q$ is measured
in MeV.

\begin{figure}[h]
\includegraphics[width=1.0\linewidth]{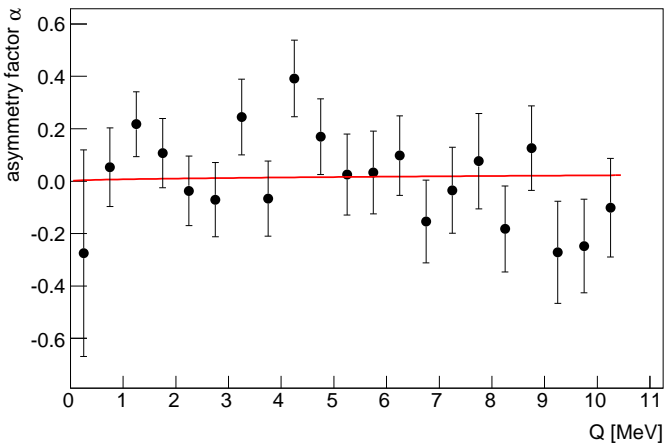}
\caption{The variation of the asymmetry factor $\alpha$ of Eq.~(\ref{alpha})
with $Q$ and fitted as a function of the $^3$He c.m.\ momentum.}
\label{asymmetry}
\end{figure}

It was shown~\cite{MER2007,SMY2007} that for $Q>4$~MeV there is a significant
but linear dependence of the unpolarised \reactionb\ differential cross
section on $\cos\theta$, where $\theta$ is the c.m.\ polar angle of the
$^3$He with respect to the deuteron beam direction. The statistics are, of
course, much higher for the unpolarised cross section than for the polarised
data presented here because of the greater strength of the unpolarised beam,
the low values of $t_{20}$ and $P_{zz}$, and the fact that the data are
spread over four polarisation modes. In order to quantify the angular
dependence of $t_{20}$, we have determined the asymmetry in $\cos\theta$ for
each energy bin by defining an asymmetry parameter through
\begin{equation}
\label{alpha}
	\alpha = {\text{d} t_{20}}/{\text{d} \cos\theta} |_{\cos\theta = 0} .
\end{equation}
The resulting values of $\alpha$ are shown versus $Q$ in
Fig.~\ref{asymmetry}.

The asymmetry parameter reflects interferences between odd and even $\eta$
partial waves so that it is an odd function of the $\eta$ c.m.\ momentum
$p_{\eta}$ which must vanish at threshold. The data shown in
Fig.~\ref{asymmetry} are compatible with zero for all $Q$ and the best fit
that is linear in $p_{\eta}$ is given by $\alpha = (0.0002 \pm 0.0005)
p_{\eta}$, where $p_{\eta}$ is measured in MeV/$c$. Only statistical errors
are quoted here and $\chi^2/\textrm{NDF} = 1.19$. There is therefore no sign
of any $s$-$p$ interference, despite the unpolarised cross section showing a
significant non-isotropy~\cite{WIL2007}.

At low energies, or at arbitrary energies in the forward and backward
directions, the spin dependence of the production amplitude can be written in
terms of two independent scalar amplitudes $A$ and $B$~\cite{GER1988}:
\begin{equation}
\label{GW}
f = \bar{u}_{^{3}\text{He}}\,\hat{\vec{p}}_p \cdot \left(A\,
\vec{\varepsilon}_{d} + i B\, \vec{\varepsilon}_d \times
\vec{\sigma}\right) u_p,
\end{equation}
where ${u}_{^{3}\text{He}}$ and $u_p$ are Pauli spinors,
$\vec{\varepsilon}_d$ is the polarisation vector of the deuteron, and
$\hat{\vec{p}}_p $ the direction of the incident proton beam in the c.m.\
frame. The differential cross section and the deuteron tensor analysing power
may then be written as:
\begin{equation}
\label{ampsquared}
\frac{\text{d}\sigma}{\text{d}\Omega} = \frac{p_{\eta}}{3p_p}\left(|A|^2+2|B|^2\right)\ \ \textrm{and}\ \
t_{20}=\sqrt{2}\,\frac{|B|^2-|A|^2}{2|B|^2+|A|^2}\cdot
\end{equation}

\begin{figure}[h]
\includegraphics[width=1.0\linewidth]{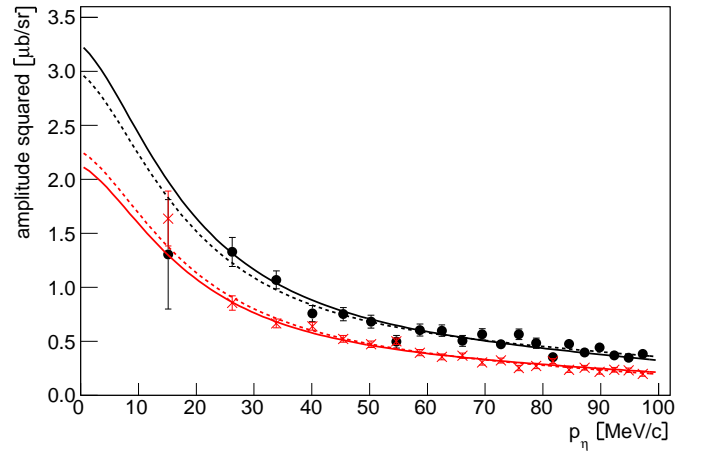}
\caption{Angular averages of $|A|^2$ (black circles) and $|B|^2$ (red
crosses) in $0.5$~MeV bins in $Q$ deduced from the current $t_{20}$
measurements and the previous ANKE unpolarised cross section
results~\cite{MER2007}. The solid lines assume that $t_{20}$ has the constant
value of $-0.21$, whereas the dashed ones include possible effects from the
slope in $Q$ given by Eq.~(\ref{bestfit}).} \label{ABsquared}
\end{figure}

Though the slope in the differential cross section~\cite{MER2007,SMY2007}
shows that the amplitudes contain small contributions from $p$ waves for
$Q\gtrsim 4$~MeV, these will not significantly affect the extrapolation to
the pole and so it is still useful to evaluate the average values of $|A|^2$
and $|B|^2$ as functions of $Q$, and this is done in Fig.~\ref{ABsquared}.
For this purpose the parameterisation of the unpolarised \reactionb\ cross
section has been taken from our earlier measurements~\cite{MER2007}.

Clearly, if the angular average of $t_{20}$ remains energy-independent, the
same has to be true for the ratio of $|A|^2/|B|^2$, despite both amplitudes
being subject to the strong FSI that leads to the violent behaviour shown in
Fig.~\ref{ABsquared}. The linear fit in Eq.~(\ref{bestfit}) gives
\begin{equation}
|B|^2/|A|^2 = (0.75\pm0.06) - (0.014\pm0.014)Q.
\label{Qratio}
\end{equation}
If such a variation exists then it is likely to be on an energy scale of
perhaps $0.75/0.014 \approx 50$~MeV rather than the less than 1~MeV
associated with the $\eta^3$He final state interaction. Initial state
interactions or the reaction mechanism itself might lead to changes on this
scale but one would need much firmer data before speculating further.

In general six invariant amplitudes are required to describe the \reactionb\
reaction~\cite{UZI2008} and an unambiguous amplitude analysis would require
very complex polarisation measurements. Attempts were made in
Ref.~\cite{WIL2007} to extend the description of the unpolarised cross
section given by Eq.~(\ref{GW}) to include the effects of $p$ waves. For this
purpose only two of the five possible $p$-wave structures were retained. In
the absence of reliable $it_{11}$ data the decomposition is very ambiguous
and the main features of our data could be well described by the ansatz:
\begin{eqnarray}
\nonumber
A&=&A_0 \left[\text{FSI}(p_{\eta}) + \alpha p_\eta\cos\theta + \beta p_{\eta}^2(3\cos^2\theta-1)/2\right],\\
B&=&B_0 \left[\text{FSI}(p_{\eta}) + \alpha p_\eta\cos\theta + \beta p_{\eta}^2(3\cos^2\theta-1)/2\right],
\label{ansatz}
\end{eqnarray}
where the FSI factor only influences the $s$-wave term. By taking $B$ to be
proportional to $A$ in terms of both $p_{\eta}$ and $\theta$, it follows
immediately that $t_{20}$ would be independent of both variables despite the
differential cross section developing a significant anisotropy at high
$p_{\eta}$~\cite{MER2007,SMY2007}. The observed linearity of the cross
section with $\cos\theta$~\cite{MER2007,SMY2007} could then arise from a
cancelation of the interference between the $\eta$ $s$ and $d$ waves and the
square of the $p$ waves.

In the $A+B$ model the vector analysing power vanishes but there is the
possibility of a non-zero $it_{11}$ in the extended model where two more spin
amplitudes $C$ and $D$ are introduced~\cite{WIL2007}. Our result that
$|it_{11}| \lesssim 0.04$ is clearly not in contradiction with the $A+B$
hypothesis outlined here.

In summary, we have measured the tensor analysing power $t_{20}$ for the
\reactionc\ reaction in 0.5~MeV steps in excess energy. The near constancy of
the angle-summed data support strongly the belief that the rapid variation of
the amplitudes with energy near threshold is due to an $s$-wave final state
interaction that is common to the two different spin states in the entrance
channel. This is consistent with the fact that fits to the unpolarised cross
section data~\cite{MER2007} yield a pole at such a low value of $|Q|$. This
would not have been the case if the amplitudes $A$ and $B$ had poles at very
different values of $Q$.

\newpage

The lack of any obvious forward/backward asymmetry in $t_{20}$, which
contrasts strongly with that observed for the cross section, gives
information on the spin structure of the scattering amplitude. Our data are
consistent with the assumption that only the amplitudes $A$ and $B$ are
important and that they have similar angular dependence. This might help in
the development of models for the \reactionb\ reaction.

We are grateful to other members of the ANKE Collaboration for their help
with this experiment and to the COSY crew for providing such good working
conditions, especially for the polarised beam. This work has been supported
in part by the JCHP FEE.

%
%

\end{document}